\documentclass[aps,preprint,showpacs,floatfix]{revtex4-1}
\usepackage{bm}
\usepackage{amsmath}
\usepackage{epsfig}
\usepackage{rotating}
\usepackage{graphicx}

\def\be{\begin{equation}}
\def\lan{\left\langle}
\def\ran{\right\rangle}
\def\ee{\end{equation}}
\def\barr{\begin{array}}
\def\earr{\end{array}}

\def\l{\left}
\def\r{\right}
\def\dis{\displaystyle}
\def\ed{\end{document}}
\def\f{\frac}

\def\co{{\cal O}}

\def\can{{\cal N}}

\def\spin{\frac{1}{2}}

\def\ed{\end{document}}
\begin{document}

\title{Multiple multi-orbit pairing algebras in shell model and interacting
boson models}

\author{V.K.B. Kota\footnote{
Phone:+917926314939, Fax:+917926314460  \\ {\it E-mail address:}
vkbkota@prl.res.in (V.K.B. Kota)}}

\affiliation{Physical Research  Laboratory, Ahmedabad 380 009, India}

\begin{abstract}

{\footnotesize{In nuclei with valence nucleons are say identical nucleons and say these
nucleons occupy several-$j$ orbits, then it is possible to consider pair
creation operator $S_+$ to be a sum of the single-$j$ shell pair creation
operators $S_+(j)$ with arbitrary phases, $S_+=\sum_j \alpha_j S_+(j);
\alpha_j=\pm 1$. In this situation, it is possible to define multi-orbit or
generalized seniority that corresponds to the quasi-spin $SU(2)$ algebra
generated by $S_+$, $S_-=(S_+)^\dagger$ and $S_0=(\hat{n} -\Omega)/2$ operators;
$\hat{n}$ is number operator and $\Omega=[\sum_j (2j+1)]/2$. There are now
multiple pairing quasi-spin $SU(2)$ algebras, one for each choice of
$\alpha_j$'s. Clearly, with $r$ number of $j$ shells there will be $2^{r-1}$ 
quasi-spin $SU(2)$ algebras. Also, the $\alpha_j$'s and the generators of the
corresponding generalized seniority generating sympletic algebras $Sp(2\Omega)$
in $U(2\Omega) \supset Sp(2\Omega)$ have one-to-one correspondence. Using these,
derived is the condition that a general one-body operator of angular momentum
rank $k$ to be a quasi-spin scalar or a vector vis-a-vis the $\alpha_j$'s. These
then will give special seniority selection rules for electromagnetic
transitions. A particular choice for $\alpha_j$'s as advocated by Arvieu and
Moszkowski (AM), based on SDI interaction, when applied to these conditions will
give the selection rules discussed in detail in the past by Talmi. We found,
using the correlation coefficient defined in the spectral distribution method of
French, that the  $\alpha_j$ choice of AM  gives pairing Hamiltonians having
maximum correlation with well known effective interactions. The various results
derived for identical fermion systems are shown to extend to identical boson
systems with the bosons occupying several-$\ell$ orbits as for example in $sd$,
$sp$, $sdg$ and $sdpf$ IBM's. The quasi-spin algebra here is $SU(1,1)$ and the
generalized seniority quantum number is generated by $SO(2\Omega)$ in
$U(2\Omega) \supset SO(2\Omega)$. The different pairing $SO(2\Omega)$ algebras
in the interacting boson models along with the tensorial nature of $E2$ and $E1$
operators in these models with respect to the corresponding $SU(1,1)$ are
presented. These different $SO(2\Omega)$ algebras will be important in the 
study of quantum phase transitions and order-chaos transitions in nuclei. }}

\end{abstract}


\maketitle

\section{Introduction}

Pairing force and the related quasi-spin or seniority quantum number continue to
play an important role in shell model in particular and nuclear structure in
general \cite{Talmi,Poves}. There are several single-$j$ shell nuclei that are
known to carry seniority quantum number as a good or useful quantum number
\cite{Talmi,Parikh,Piet}. Even when single shell seniority is a broken symmetry,
seniority quantum number provides a basis for constructing shell model
Hamiltonian matrices \cite{French}. Pairing symmetry with nucleons occupying
several $j$-orbits is more complex and less well understood from the point of
view of its goodness or usefulness in nuclei. Restricting to nuclei with valence
nucleons are identical nucleons (protons or neutrons), and say these nucleons
occupy several-$j$ orbits, then it is possible to consider pair creation
operator $S_+$ to be a sum of the single-$j$ shell pair creation operators
$S_+(j)$ with arbitrary phases, $S_+=\sum_j \alpha_j S_+(j); \alpha_j=\pm 1$. In
this situation, it is possible to define multi-orbit or generalized seniority
that corresponds to the quasi-spin $SU(2)$ algebra. However, with $r$ number of
$j$ shells there will be $2^{r-1}$  quasi-spin $SU(2)$ algebras. Also, the
$\alpha_j$'s and the generators of the corresponding generalized seniority
generating sympletic algebras $Sp(2\Omega)$ in $U(2\Omega) \supset Sp(2\Omega)$
have one-to-one correspondence. In this paper we will examine in detail these
multiple pairing SU(2) algebras and also the corresponding multiple pairing
algebras for interacting boson systems. The usefulness or goodness of these
multiple pairing algebras is not well known except a special situation was
studied long time back by Arvieu and Moszkowski (AM) \cite{Arvieu} in the
context of surface delta interaction. In addition, pair states with $\alpha_j$
being free parameters (need not be $+1$ or $-1$) are used in generating
low-lying states with good  generalized seniority \cite{Talmi} and they are also
employed in the so called  broken pair model \cite{Gambhir}. On the other hand
these are also used in  providing a microscopic basis for the interacting boson
model \cite{Iac-Tal}.  Going beyond all these, there are also attempts to solve
and apply more  general pairing Hamiltonian's by Pan Feng et al \cite{pfeng} and
also a pair shell model is being studied by Zhao et al \cite{zhao}. Now we will
give a preview.

Section II gives in some detail the algebraic structure of the multiple
multi-orbit pairing quasi-spin $SU(2)$ and the complimentary $Sp(N)$ algebras in
$j-j$ coupling shell model for identical nucleons. Using the multiple algebras,
in Section III derived are the selection rules for electromagnetic transitions
with multi-orbit seniority. In section IV, correlation between realistic
effective interactions and pairing operator with a given set of phases
($\alpha_j$) is studied and shown that the choice advocated by AM gives maximum
correlation. Section V gives details of the algebraic structure of the multiple
multi-orbit pairing quasi-spin $SU(1,1)$ and the complimentary $SO(N)$ algebras
in interacting boson models with identical bosons such as $sd$, $sp$, $sdg$ and
$sdpf$ IBM's. Here, again derived are the selection rules for electromagnetic
transition operators as a function of the given set of phases in the generalized
boson pair operator. In Section VI presented are the results for the particle
number dependence of the matrix elements of one-body operators that are
quasi-spin scalar or vector for both fermion and boson systems. In Section VII
presented are some applications of the multi-orbit pairing algebras in shell
model and interacting boson models. Finally, Section VIII gives conclusions and
future outlook.

\section{Multiple multi-orbit pairing quasi-spin $SU(2)$ and the complimentary
$Sp(N)$ algebras in $j-j$ coupling shell model}

\subsection{Multiple multi-orbit pairing quasi-spin $SU(2)$ algebras}
  
Let us say there are $m$ number of identical fermions (protons or neutrons) in
$j$ orbits $j_1$, $j_2$, $\ldots$, $j_r$. Now, it is possible to define a
generalized pair creation operator $S_+$ as
\be
S_+ = \dis\sum_{j} \alpha_j S_+(j)\;;\;S_+(j)=\dis\sum_{m>0} (-1)^{j-m} 
a^\dagger_{jm} a^\dagger_{j-m} = \dis\frac{\dis\sqrt{2j+1}}{2}\;
\l(a^\dagger_j a^\dagger_j\r)^0\;.
\label{eq1}
\ee
Here, $\alpha_j$ are free parameters and assumed to be real. The $m$ used for
number of particles should not be confused with the $m$ in $jm$. Given the $S_+$
operator, the corresponding pair annihilation operator $S_-$ is
\be
S_- = \l(S_+\r)^\dagger = \dis\sum_{j} \alpha_j S_-(j)\;;\;S_-(j)
=\l(S_+(j)\r)^\dagger = -\dis\frac{\dis\sqrt{2j+1}}{2}\;\l(\tilde{a}_j 
\tilde{a}_j\r)^0\;.
\label{eq2}
\ee
Note that $a_{jm}=(-1)^{j-m} \tilde{a}_{j-m}$. The operators $S_+$, $S_-$ and 
$S_0$, with $\hat{n} = \sum_{jm} a^\dagger_{jm} a_{jm}$ the number operator,
\be
S_0=\dis\frac{\hat{n} - \Omega}{2}\;;\;\Omega=\sum_j \Omega_j\;,\;\;
\Omega_j=(2j+1)/2\;.
\label{eq3}
\ee
form the generalized quasi-spin SU(2) algebra [hereafter called $SU_Q(2)$] 
only if
\be
\alpha_j^2=1\;\mbox{for all}\;\;j\;.
\label{eq4}
\ee
With Eq. (\ref{eq4}) we have,
\be
\l[S_0\;S_\pm\r]=\pm S_\pm\;,\;\;\l[S_+\;S_-\r]=2S_0\;.
\label{eq4a}
\ee
Thus, in the multi-orbit situation for each 
$$
\l\{\alpha_{j_1}, \alpha_{j_2}, \ldots, \alpha_{j_r}\r\}
$$
with $\alpha_{j_i}=\pm 1$ there is a $SU_Q(2)$ algebra defined by the operators
in Eqs. (\ref{eq1}), (\ref{eq2}) and (\ref{eq3}). For example, say we have three
$j$ orbits $j_1$, $j_2$ and $j_3$. Then, without loss of generality we can
choose $\alpha_1=+1$ and then $(\alpha_2,\alpha_3)$ can take values $(+1,+1)$,
$(+1,-1)$, $(-1,+1)$, $(-1,-1)$ giving four pairing $SU_Q(2)$ algebras.
Similarly, with four $j$ orbits, there will be eight $SU_Q(2)$ algebras and in
general for $r$ number of $j$ orbits there will $2^{r-1}$ number of $SU_Q(2)$
algebras. The consequences of having these multiple pairing $SU_Q(2)$ algebras
will be investigated in the following.

Though well known, for later use and for completeness, some of the results of
the $SU_Q(2)$ algebra are that the  $S^2 = S_+ S_- -  S_0 + S_0^2$ operator and
the $S_0$ operator in Eq. (\ref{eq3}) define the quasi-spin $s$ and its
$z$-component $m_s$ with  $S^2\l| sm_s \ran = s(s+1) \l| s m_s \ran$ and
$S_0 \l|sm_s \ran = m_s \l| sm_s \ran$. Also, from Eq. (\ref{eq3}) we have
$m_s=(m-\Omega)/2$; the $m$ here is number of particles. Moreover, it is
possible to introduce the so called seniority quantum number $v$ such that
$s=(\Omega-v)/2$ giving,
\be
\barr{rcl}
s&=&\l( \Omega - v \r)/2\;\;,m_s = \l( m-\Omega \r)/2\;,\\
v &=& m, m-2, \ldots, 0 \;\mbox{ or }\; 1\; \mbox{ for }\; m \; \leq \Omega\\
&=& \l(2\Omega -m \r), \l(2\Omega-m \r)-2, ..,\; 0\; \mbox{ or }\; 1
\; \mbox{ for } \;  m \geq \Omega \;\; .
\earr \label{eq5}
\ee
Note that the total number of single particle states is $N=2\Omega$ and 
therefore for $m > \Omega$ one has fermion holes rather than particles. The 
following results will provide a meaning to the seniority quantum number 
``$v$'',
\be
\barr{rll}
{\lan S_+ S_- \ran}^{sm_s} &=& = 
{\lan S_+ S_- \ran}^{mv} = {\lan mv \mid S_+ S_- \mid mv
\ran} \\
&=& \frac{1}{4} (m-v) ( 2 \Omega-m-v+2) \;,
\earr \label{eq6}
\ee
\be
\l|\l. m,v, \beta \ran\r. =
\sqrt{\f{\dis \l(\Omega-v-p\r)!}{\dis \l(\Omega-v\r)! p!}} \l(S_+
\r)^{\f{\dis m-v}{\dis 2}} \l|\l. v, v, \beta \ran\r. ;\;
p=\f{(m-v)}{2} \; .
\label{eq7}
\ee
With these, it is clear that for a given $v$ and $m$ there are $(m-v)/2$ zero
coupled pairs. Thus, $v$ gives the number of particles that are not coupled to
angular momentum zero. In Eq. (\ref{eq7}), $\beta$ is an extra label that is
required to specify a $(j_1,j_2,\ldots,j_r)^m$ state completely.

Before going further, an important result (to be used later) that follows from
Eqs. (\ref{eq1}) and (\ref{eq2}) is,
\be
\barr{rcl}
4S_+S_- & = & 4\dis\sum_jS_+(j) S_-(j) + \dis\sum_{j_1 > j_2} \alpha_{j_1} 
\alpha_{j_2} \\
& \times & \dis\sum_k \dis\sqrt{2k+1}\,\l\{\l[\l(a^\dagger_{j_1}
\tilde{a}_{j_2}\r)^k \l(a^\dagger_{j_1}\tilde{a}_{j_2}\r)^k\r]^0 +
\l[\l(a^\dagger_{j_2}\tilde{a}_{j_1}\r)^k \l(a^\dagger_{j_2}\tilde{a}_{j_1}
\r)^k\r]^0\r\}\;.
\earr \label{eq8}
\ee

\subsection{Multiple multi-orbit complimentary pairing $Sp(N)$  algebras}

In the $(j_1,j_2,\ldots,j_r)^m$ space, often it is more convenient to start with
the $U(N)$ algebra generated by the one-body operators $u^k_q(j_1,j_2)$,
\be
u^k_q(j_1,j_2) = \l(a^\dagger_{j_1}\tilde{a}_{j_2}\r)^k_q\;.
\label{eq9}
\ee
Total number of generators is obviously $N^2$ and $N=2\Omega$. All $m$ fermion 
states will be antisymmetric and therefore belong uniquely to the irreducible 
representation (irrep) $\{1^m\}$ of $U(N)$. The quadratic Casimir invariant of 
$U(N)$ is easily given by
\be
C_2(U(N))=\dis\sum_{j_1,j_2} (-1)^{j_1-j_2} \dis\sum_k 
u^k(j_1,j_2)\cdot u^k(j_2,j_1) \;,
\label{eq10}
\ee
with eigenvalues
\be
\lan C_2(U(N))\ran^m = m(N+1-m)\;;\;\;N=2\Omega\;.
\label{eq11}
\ee
Eq. (\ref{eq11}) can be proved by writing the one and two-body parts of 
$C_2(U(N))$ and then showing that the one-body part is $2\Omega \hat{n}$ and 
the two-body part will have two-particle matrix elements diagonal with all of 
them having value $-2$.

More importantly, $U(N) \supset Sp(N)$ and the $Sp(N)$ algebra is generated by
the $N(N+1)/2$ number of generators $u^k_q(j,j)$ with $k$=odd only and
$V^k_q(j_1,j_2)$, $j_1 > j_2$ with 
\be
V^k_q(j_1,j_2)=\l[{\cal N}(j_1,j_2,k)\r]^{1/2} \l[\l(a^\dagger_{j_1}
\tilde{a}_{j_2}\r)^k_q + X(j_1,j_2,k)\,\l(a^\dagger_{j_2}
\tilde{a}_{j_1}\r)^k_q\r]\;,\;\;\{X(j_1,j_2,k)\}^2=1\;.
\label{eq12}
\ee
The quadratic Casimir invariant of $Sp(N)$ is given by,
\be
C_2(Sp(N)) = 2 \dis\sum_{j} \dis\sum_{k=odd} u^k(j,j) \cdot u^k(j,j) + 
\dis\sum_{j_1 > j_2;k} V^k(j_1,j_2) \cdot V^k(j_1,j_2)\;.
\label{eq13}
\ee
The $Sp(N)$ algebra will be complimentary to the quasi-spin $SU(2)$ algebra 
defined for a given set of $\l\{\alpha_{j_1}, \alpha_{j_2}, \ldots, 
\alpha_{j_r}\r\}$ provided
\be
{\cal N}(j_1,j_2,k)=(-1)^{k+1} \alpha_{j_1} \alpha_{j_2}\;,\;\;\;X(j_1,j_2,k)
=(-1)^{j_1 + j_2 +k} \alpha_{j_1} \alpha_{j_2}\;.
\label{eq14}
\ee
Using Eqs. (\ref{eq10}) and (\ref{eq12})-(\ref{eq14}) along with Eq. (\ref{eq8})
it is easy to derive the following important relation,
\be
C_2(U(N)) - C_2(Sp(N)) = 4S_+S_- - \hat{n}\;.
\label{eq15}
\ee
Now, Eqs. (\ref{eq15}), (\ref{eq11}) and (\ref{eq6}) will give
\be
\lan C_2(Sp(N) \ran^{m,v} = v(2\Omega+2-v)
\label{eq16})
\ee
and this proves that the seniority quantum number $v$ corresponds to the 
$Sp(N)$ irrep $\lan 1^v\ran$. 

In summary, given the $SU_Q(2)$ algebra generated by $\{S_+, S_- , S_0\}$
operators for a given set of  $\l\{\alpha_{j_1}, \alpha_{j_2}, \ldots,
\alpha_{j_r}\r\}$ with $\alpha_{j_i}=+1$ or $-1$, there is a complimentary
($\leftrightarrow$) $Sp(N)$ subalgebra of $U(N)$ generated by
\be
\barr{l}
Sp(N)\,:\, u^k(j,j)=\l(a^\dagger_{j}\tilde{a}_{j}\r)^k_q\;\;
\mbox{with}\;\; k=\mbox{odd}\;, \\
V^k_q(j_1,j_2)=\l[(-1)^{k+1} \alpha_{j_1} \alpha_{j_2}\r]^{1/2} 
\l[\l(a^\dagger_{j_1}\tilde{a}_{j_2}\r)^k_q + (-1)^{j_1+j_2+k} \alpha_{j_1} 
\alpha_{j_2}\,\l(a^\dagger_{j_2}\tilde{a}_{j_1}\r)^k_q\r]\;\;\mbox{with}
\;\;j_1 > j_2\;.
\earr \label{eq17}
\ee
As the $Sp(N)$ generators are one-body operators and that $Sp(N) \leftrightarrow
SU_Q(2)$, there will be special selection rules for electro-magnetic transition
operators connecting $m$ fermion states with good seniority. Though these are
well known for a special choice of $\alpha$'s \cite{Talmi}, their relation to
the multiple $SU(2)$ algebras or equivalently to the $\l\{\alpha_{j_1},
\alpha_{j_2}, \ldots, \alpha_{j_r}\r\}$ set is, to our best of knowledge, is not
discussed before. We will turn to this now.

\section{Selection rules for electro-magnetic transitions with multi-orbit
seniority}

Electro-magnetic operators are essentially one-body operators (two and
higher-body terms are usually not considered). In order to derive selection
rules and matrix elements for allowed  transitions, let us first consider the
commutator of $S_+$ with $\l(a^\dagger_{j_1}\tilde{a}_{j_2}\r)^k_q$. Firstly we
have easily,
\be
\l[S_+(j)\;,\;\l(a^\dagger_{j_1}\tilde{a}_{j_2}\r)^k_q \r] =
-\delta_{j,j_2}\; \l(a^\dagger_{j_1} a^\dagger_{j_2}\r)^k_q\;.
\label{eq18}
\ee
This gives
\be\barr{l}
\l[S_+\;,\;\l(a^\dagger_{j_1}\tilde{a}_{j_2}\r)^k_q + X
\l(a^\dagger_{j_2} \tilde{a}_{j_1}\r)^k_q \r] \\
=-\alpha_{j_2}\,\l(a^\dagger_{j_1} a^\dagger_{j_2}\r)^k_q \;
\l\{1-X\,\alpha_{j_1}\,\alpha_{j_2}\,(-1)^{j_1+j_2+k}\r\} \\
=0\;\;\mbox{if}\;\;X=\alpha_{j_1}\,\alpha_{j_2}\,(-1)^{j_1+j_2+k} \\
\neq 0 \;\;\mbox{if}\;\;X=-\alpha_{j_1}\,\alpha_{j_2}\,(-1)^{j_1+j_2+k} \;.
\earr \label{eq19}
\ee
Note that the commutator is zero implies that the operator is a scalar $T^0_0$
with respect to $SU_Q(2)$ and otherwise it will be a quasi-spin vector $T^1_0$.
In either situation the $S_z$ component of $T$ is zero as a one-body operator
can not change particle number. Thus, for $j_1 \neq j_2$ we have
\be
\barr{l}
U^k_q(j_1,j_2)={\cal N}_u\;\l\{\l(a^\dagger_{j_1}\tilde{a}_{j_2}\r)^k_q + 
\alpha_{j_1}\,\alpha_{j_2}\,(-1)^{j_1+j_2+k} 
\l(a^\dagger_{j_2}\tilde{a}_{j_1}\r)^k_q\r\} \rightarrow T^0_0 \;,\\
W^k_q(j_1,j_2)={\cal N}_w\;\l\{\l(a^\dagger_{j_1}\tilde{a}_{j_2}\r)^k_q -
\alpha_{j_1}\,\alpha_{j_2}\,(-1)^{j_1+j_2+k} 
\l(a^\dagger_{j_2}\tilde{a}_{j_1}\r)^k_q\r\} \rightarrow T^1_0 \;.
\earr \label{eq20}
\ee
Here ${\cal N}_u$ and ${\cal N}_w$ are some constants. Similarly, for $j_1=j_2$
we have
\be
\barr{l}
\l(a^\dagger_{j}\tilde{a}_{j}\r)^k_q \;\;\mbox{with}\;\;k\;\;\mbox{odd} \;
\rightarrow T^0_0 \;,\\
\l(a^\dagger_{j}\tilde{a}_{j}\r)^k_q \;\;\mbox{with}\;\;k\;\;\mbox{even} \;
\rightarrow T^1_0 \;\;\;\mbox{except for}\;\;k=0\;.
\earr \label{eq21}
\ee
The results in Eqs. (\ref{eq20}) are easy to understand as $U^k_q$ in Eq.
(\ref{eq20}) is to within a factor same as $V^k_q$ of Eq. (\ref{eq17}) and
therefore a generator of $Sp(N)$. Hence it can not change the $v$ quantum number
of a $m$-particle state. Also, as $Sp(N) \leftrightarrow SU_Q(2)$,  clearly
$U^k_q$ will be a $SU_Q(2)$ scalar. Similarly turning to Eq. (\ref{eq21}), as 
$\l(a^\dagger_{j}\tilde{a}_{j}\r)^k_q$ with $k$ odd are generators of $Sp(N)$
and hence they are also $SU_Q(2)$ scalars. 

General form of electric and magnetic multipole operators $T^{EL}$ and $T^{ML}$
respectively with $L=1,2,3,\ldots$ is, with $X=E$ or $M$,
\be
\barr{rcl}
T^{XL}_q & = & \dis\sum_{j_1,j_2} \epsilon^{XL}_{j_1,j_2} \l(a^\dagger_{j_1}
\tilde{a}_{j_2}\r)^L_q \\
& = & \dis\sum_j \epsilon^{XL}_{j,j}\l(a^\dagger_{j}\tilde{a}_{j}\r)^L_q
+\dis\sum_{j_1 > j_2} \epsilon^{XL}_{j_1,j_2} \l[
\l(a^\dagger_{j_1}\tilde{a}_{j_2}\r)^L_q +\dis\frac{\epsilon^{XL}_{j_2,j_1}}
{\epsilon^{XL}_{j_1,j_2}}
\l(a^\dagger_{j_2}\tilde{a}_{j_1}\r)^L_q\r]\;.
\earr \label{eq22}
\ee
Therefore, $\epsilon^{XL}_{j_2,j_1} / \epsilon^{XL}_{j_1,j_2}$ along with
Eqs. (\ref{eq20})and (\ref{eq21}) will determine the selection rules. Then,
\be
\barr{l}
\dis\frac{\epsilon^{XL}_{j_2,j_1}}{\epsilon^{XL}_{j_1,j_2}}
=\alpha_{j_1} \alpha_{j_2} (-1)^{j_1+j_2+L} \rightarrow T^0_0\;\;
\mbox{w.r.t.}\;\;SU_Q(2)\;, \\
\dis\frac{\epsilon^{XL}_{j_2,j_1}}{\epsilon^{XL}_{j_1,j_2}}
=-\alpha_{j_1} \alpha_{j_2} (-1)^{j_1+j_2+L} \rightarrow T^1_0\;\;
\mbox{w.r.t.}\;\;SU_Q(2)\;.
\earr \label{eq22a}
\ee
Thus, the $SU_Q(2)$ tensorial nature of $T^{XL}$ depend on the $\alpha_i$
choice. For $T^0_0$ we have $v \rightarrow v$ and for $T^1_0$ we have $v
\rightarrow v,\,v \pm 2$ transitions. It is well known \cite{Talmi,Arvieu} that
for $T^{EL}$ and  $T^{ML}$ operators,
\be
\dis\frac{\epsilon^{EL}_{j_2,j_1}}{\epsilon^{EL}_{j_1,j_2}} =
-(-1)^{\ell_1+\ell_2+j_1+j_2+L} \;\;,\;\;\;\;\;
\dis\frac{\epsilon^{ML}_{j_2,j_1}}{\epsilon^{ML}_{j_1,j_2}} =
(-1)^{\ell_1+\ell_2+j_1+j_2+L}\;.
\label{eq23}
\ee
In Eq. (\ref{eq23}) $\ell_i$ is the orbital angular momentum of the $j_i$ orbit.
Therefore, combining  results in Eqs. (\ref{eq20})-(\ref{eq23}) together with
parity selection rule will give seniority selection rules, in the multi-orbit
situation, for electro-magnetic transition operators when the observed states
carry seniority quantum number as a good quantum number. The selection rules
with the choice $\alpha_{j_i}=(-1)^{\ell_i}$ for all $i$ are as follows.

\begin{enumerate}

\item $T^{EL}$ with $L$ even will be $T^1_0$ w.r.t. $SU_Q(2)$.

\item $T^{EL}$ with $L$ odd will be $T^1_0$ w.r.t. $SU_Q(2)$. However, if all 
$j$ orbits have same parity, then $T^{EL}$ with $L$ odd will not exist. 
Therefore here, for the transitions to occur, we need minimum two orbits of 
different parity.

\item $T^{ML}$ with $L$ odd will be $T^0_0$ w.r.t. $SU_Q(2)$.

\item $T^{ML}$ with $L$ even will be $T^0_0$ w.r.t. $SU_Q(2)$. However, if all 
$j$ orbits have same parity, then $T^{ML}$ with $L$ even will not exist. 
Therefore here, for the transitions to occur, we need minimum two orbits of 
different parity.

\item For $T^0_0$ only $v \rightarrow v$ transitions are allowed while for
$T^1_0$ both $v \rightarrow v$ and $v \rightarrow v \pm 2$ transition are
allowed. For both $m$ is not changed.

\end{enumerate}

\noindent The above rules were given already by AM \cite{Arvieu} and described
by Talmi \cite{Talmi}. As stated by Arvieu and Moszkowski, they have introduced
the choice $\alpha_i=(-1)^{\ell_i}$ \lq\lq{for convenience }\rq\rq and then
found that it will make surface delta interaction a $SU_Q(2)$ scalar. It is
important to note that for $SU_Q(2)$ generated by $\alpha_i \neq (-1)^{\ell_i}$,
the above rules (1)-(4) will be violated and then Eq. (\ref{eq22a}) has to be
applied. This is a new result not reported before, to our knowledge, in the
literature. A similar result applies to interacting boson models as presented
ahead in Section V. Before going further, within shell model context it is
necessary to conform that a realistic pairing operator do respect the condition
$\alpha_i=(-1)^{\ell_i}$. In order to test this, we will use correlation
coefficient between operators as defined in French's spectral distribution
method \cite{KH-10}.

\section{Correlation between operators and phase choice in the pairing
operator}

Given an operator $\co$ acting in $m$ particle spaces and assumed to be real,
its $m$ particle trace is $\lan\lan \co \ran\ran^m = \sum_\alpha\,\lan m, \alpha
\mid \co \mid m,\alpha\ran$ where $\l.\l| m,\alpha\r.\ran$ are $m$-particle
states. Similarly, the $m$-particle average is $\lan \co\ran^m =[d(m)]^{-1}
\lan\lan \co \ran\ran^m$ where $d(m)$ is $m$-particle space dimension. In $m$
particle spaces it is possible to define, using the spectral distribution method
of French \cite{CFT,KH-10}, a geometry \cite{CFT,Potbhare} with norm (or size or
length) of an operator $\co$ given by $\mid\mid \co \mid\mid_m = \sqrt{\lan
\tilde{\co}\tilde{\co} \ran^m}$; $\tilde{\co}$ is the traceless part of $\co$.
With this, given two operators $\co_1$ and $\co_2$, the correlation
coefficient  
\be
\zeta(\co_1,\co_2)= \dis\frac{\lan \widetilde{\co_1} \widetilde{\co_2} \ran^m}{
\mid\mid \co_1 \mid\mid_m \;\mid\mid \co_2 \mid\mid_m}\;,
\label{eqss1}
\ee
gives the cosine of the angle between the two operators. Thus, $\co_1$ and
$\co_2$ are same within a normalization constant if $\zeta=1$ and they are
orthogonal to each other if $\zeta=0$ \cite{Potbhare,KH-10}. Most recent
application of norms and correlation coefficients defined above to understand
the structure of effective interactions is due to Draayer et al
\cite{Draayer1,Draayer2}.

Clearly, in a given shell model space, given a realistic effective interaction
Hamiltonian $H$, the $\zeta$ in Eq. (\ref{eqss1}) can be used as a measure for
its closeness to the pairing Hamiltonian $H_P=S_+S_-$ with $S_+$ defined by Eq.
(\ref{eq1}) for a given set of $\alpha_j$'s. Evaluating $\zeta(H,H_P)$ for all
possible $\alpha_j$ sets, it is possible to identify the $\alpha_j$ set that
gives maximum correlation of $H_p$ with $H$. Following this, $\zeta(H,H_P)$ is
evaluated for effective interactions in ($^0f_{7/2}$, $^0f_{5/2}$, $^1p_{3/2}$,
$^1p_{1/2}$), ($^0f_{5/2}$, $^1p_{3/2}$, $^1p_{1/2}$, $^0g_{9/2}$) and
($^0g_{7/2}$, $^1d_{5/2}$, $^1d_{3/2}$, $^2s_{1/2}$, $^0h_{11/2}$) spaces using
GXPF1 \cite{gxpf1}, JUN45 \cite{jun45} and jj55-SVD \cite{jj55} interactions
respectively. As we are considering only identical particle systems and also as
we are interested in studying the correlation of $H$'s with $H_P$'s, only the
$T=1$ part of the interactions is considered (dropped are the $T=0$ two-body
matrix elements and also the single particle energies). With this $\zeta(H,H_P)$
are calculated in the three spaces for different values of the particle number
$m$ and for all possible choices of $\alpha_j$'s defining $S_+$ and hence $H_P$.
Results are given in Table I. It is clearly seen that the choice
$\alpha_j=(-1)^{\ell_i}$ gives the largest value for $\zeta$ and hence it should
be the most preferred choice. This is a significant result justifying the choice
made by AM \cite{Arvieu}, although the magnitude of $\zeta$ is not more than
$0.3$. Thus, realistic $H$ are far, on a global $m$-particle space scale, from
the simple pairing Hamiltonian. However, it is likely that the generalized
pairing quasi-spin or sympletic symmetry may be an effective symmetry for
low-lying state and some special high-spin states \cite{Piet}. Evidence for this
will be discussed in Section VII. 

Before turning to interacting boson systems, it is useful to add that in
principle the spectral distribution method can be used to study the mixing of
seniority quantum number in the eigenstates generated by a given Hamiltonian
by using the so called partial variances \cite{KH-10,Parikh}. The $v_i
\rightarrow v_f$ partial variances, with $v_i \neq v_f$, are defined by
\be
\sigma^2(m, v_i \rightarrow m, v_f) = \l[d(m,v_i)\r]^{-1} \dis\sum_{\alpha,
\beta} \l|\lan  m , v_f , \beta \mid H \mid m , v_i, \alpha\ran\r|^2\;.
\label{eqvmix}
\ee
In Eq. (\ref{eqvmix}), $d(m,v)$ is the dimension of the $(m,v)$ space.
It is important to note that the partial variances can be evaluated without
constructing the $H$ matrices but by using the propagation equations. These are
available both for fermion and boson systems; see \cite{Quesne,kota}. However,
propagation equations for the more realistic $\sigma^2(m, v_i, J \rightarrow m,
v_f , J)$ partial variances are not yet available.

\begin{table*}

\caption{Correlation coefficient $\zeta$ between a realistic interaction ($H$)
and the pairing Hamiltonian $H_p$ for various particle numbers ($m$) in three
different spectroscopic spaces. The single particle (sp) orbits for these three
spaces are given in column \#1. The range of $m$ values used for each sp space
is given in column \#3. The phases $\alpha_j$ for each orbit in the generalized
pair creation operator are given in column \#4 (the order is same as the sp
orbits listed in column \#1). The variation in $\zeta$ with particle number $m$
is given in column \#5. Results for the phase choices that give $|\zeta| < 0.1$
for all $m$ values are not shown in the table.  See text for other details.}

\begin{tabular}{c|c|c|c|l}
\hline 
\hline
sp orbits & interaction & $m$ & $\alpha_j$ & $\zeta(H,H_p)$ \\
\hline
\hline
$^0g_{7/2}$, $^1d_{5/2}$, $^1d_{3/2}$, $^2s_{1/2}$, 
$^0h_{11/2}$ & jj55-SVD & $2-30$& $(+,+,+,+,-)$ &  $0.33$-$0.11$ \\
& & & $(+,+,+,-,-)$ & $0.26$-$0.09$ \\
& & & $(+,+,-,+,-)$ & $0.17$-$0.06$ \\
& & & $(+,+,-,-,-)$ & $0.13$-$0.04$ \\
& & & $(+,-,+,+,-)$ & $0.11$-$0.04$ \\
$^0f_{5/2}$, $^1p_{3/2}$, $^1p_{1/2}$, $^0g_{9/2}$ & jun45 & 
$2-20$& $(+,+,+,-)$ &  $0.42$-$0.21$ \\
& & & $(+,+,-,-)$ & $0.27$-$0.13$ \\
& & & $(+,-,+,-)$ & $0.15$-$0.07$ \\
& & & $(+,-,-,-)$ & $0.12$-$0.06$ \\
$^0f_{7/2}$, $^1p_{3/2}$, $^0f_{5/2}$, $^1p_{1/2}$ & gxpf1 & 
$2-18$& $(+,+,+,+)$ &  $0.36$-$0.33$ \\
& & & $(+,+,+,-)$ & $0.22$-$0.20$ \\
& & & $(+,-,+,+)$ & $0.13$-$0.12$ \\
& & & $(+,-,+,-)$ & $0.13$-$0.11$ \\
& & & $(+,-,-,-)$ & $0.11$-$0.10$ \\
\hline\hline
\end{tabular}
\label{corr}
\end{table*}

\section{Multiple multi-orbit pairing quasi-spin $SU(1,1)$ and the complimentary
$SO(N)$ algebras in interacting boson models}

Going beyond the shell model, also within the interacting boson models, i.e. for
example in $sd$, $sp$, $sdg$ and $sdpf$ IBM's, again it is possible to have 
multiple pairing symmetry algebras as we have several $\ell$ orbits in these
models with bosons \cite{KOYD,KS-16,Ko-00,Kusnezov}. Here, as it is well known,
the pairing  algebra is $SU_Q(1,1)$ instead of $SU_Q(2)$ \cite{Ui}. Let us
consider IBM with identical bosons carrying angular momentum $\ell_1$, $\ell_2$,
\ldots, $\ell_r$ and the parity of an $\ell_i$ orbit is $(-1)^{\ell_i}$. Now,
again it is possible to define a generalized boson pair creation operator
$S^B_+$ as
\be
S^B_+ = \dis\sum_{\ell} \beta_\ell S^B_+(\ell)\;;\;S^B_+(\ell)=\spin 
\dis\sum_{m} (-1)^{m} b^\dagger_{\ell m} b^\dagger_{\ell -m} = 
\dis\frac{\dis\sqrt{2\ell +1}}{2}\;(-1)^\ell\;\l(b^\dagger_\ell b^\dagger_\ell 
\r)^0 = \spin b^\dagger_\ell \cdot b^\dagger_\ell\;.
\label{eqb1}
\ee
Here, $\beta_\ell$ are free parameters and assumed to be real.  Given the 
$S^B_+$ operator, the corresponding pair annihilation operator $S^B_-$ is
\be
S^B_- = \l(S^B_+\r)^\dagger = \dis\sum_{\ell} \beta_\ell S^B_-(\ell)\;;\;
S^B_-(\ell)
=\l(S^B_+(\ell)\r)^\dagger = (-1)^{\ell} \dis\frac{\dis\sqrt{2\ell +1}}{2}\;
\l(\tilde{b}_\ell \tilde{b}_\ell \r)^0\ = \spin \tilde{b}_\ell \cdot 
\tilde{b}_\ell \;.
\label{eqb2}
\ee
Note that $b_{\ell m}=(-1)^{l-m} \tilde{b}_{\ell -m}$. The operators $S^B_+$, 
$S^B_-$ and $S^B_0$, with $\hat{n}^B = \sum_{\ell m} a^\dagger_{\ell m} 
a_{\ell m}$ the number operator,
\be
S^B_0=\dis\frac{\hat{n}^B + \Omega^B}{2}\;;\;\Omega^B=\sum_\ell \Omega^B_\ell,
\;\;\Omega^B_\ell=(2\ell +1)/2\;.
\label{eqb3}
\ee
form the generalized quasi-spin SU(1,1) algebra [hereafter called $SU^B_Q(1,1)$]
only if
\be
\beta_\ell^2=1\;\mbox{for all}\;\;\ell\;.
\label{eqb4}
\ee
With Eq. (\ref{eqb4}) we have,
\be
\l[S^B_0\;S^B_\pm\r]=\pm S^B_\pm\;,\;\;\l[S^B_+\;S^B_-\r]=-2S^B_0\;.
\label{eqb5}
\ee
Thus, in the multi-orbit situation for each 
$$
\l\{\beta_{\ell_1}, \beta_{\ell_2}, \ldots, \beta_{\ell_r}\r\}
$$
with $\beta_{\ell_i}=\pm 1$ there is a $SU^B_Q(1,1)$ algebra defined by the
operators in Eqs. (\ref{eqb1}), (\ref{eqb2}) and (\ref{eqb3}). In general for
$r$ number of $\ell$ orbits there will $2^{r-1}$ number of $SU^B_Q(1,1)$
algebras. Let us mention that $(S^B)^2 = (S^B_0)^2 - S^B_0 - S^B_+ S^B_-$ and 
$S^B_0=(\hat{n}^B + \Omega^B)/2$ provide the quasi-spin $s$ and the $s_z$ 
quantum number $m_s$ giving the basis $\l.\l|s,m_s\r.\ran$ \cite{KOYD,KS-16},
\be
\barr{l}
(S^B)^2\l.\l|s,m_s,\gamma\r.\ran=s(s-1)\l.\l|s,m_s,\gamma\r.\ran\;,\;\;
S_0 \l.\l|s,m_s,\gamma\r.\ran=m_s\l.\l|s,m_s,\gamma\r.\ran\;; \\
m_s=s,s+1,s+2,\ldots \\
\Rightarrow \\
s=(\Omega^B+\omega^B)/2\;,\;\;m_s=(\Omega^B+N^B)/2\;,\;\;
\omega^B=N^B,N^B-2,\dots,0\;\mbox{or}\;
1,\\
S^B_+ S^B_-\l.\l|s,m_s,\gamma\r.\ran = S^B_+ S^B_-\l.\l|N^B,\omega^B,\gamma
\r.\ran = \frac{1}{4}(N^B -\omega^B)(\omega^B+N^B+2\Omega^B-2)
\l.\l|N^B,\omega^B,
\gamma\r.\ran \;. 
\earr \label{eqb6}
\ee
Here, $N^B$ is number of bosons. Just as for fermions, corresponding to each 
$SU_Q^B(1,1)$ there will be, in the $(\ell_1, \ell_2,\ldots, \ell_r)^{N^B}$ 
space, a $SO(\can)$ in $U(\can)$ with $\can=2\Omega^B=\sum_\ell (2\ell +1)$. 
The $U(\can)$ algebra is generated by the $\can^2$ number of operators 
$$
u^k_q(\ell_1 , \ell_2)=\l(b^\dagger_{\ell_1}\tilde{b}_{\ell_2}\r)^k_q\;.
$$ 
As all the $N^B$ boson states will be symmetric, they belong 
uniquely to the irrep $\{N^B\}$ of $U(\can)$. The quadratic Casimir invariant 
of $U(\can)$ is easily given by
\be
C_2(U(\can))=\dis\sum_{\ell_1,\ell_2} (-1)^{\ell_1+\ell_2} \dis\sum_k 
u^k(\ell_1, \ell_2)\cdot u^k(\ell_2, \ell_1) \;,
\label{eqb10}
\ee
with eigenvalues
\be
\lan C_2(U(\can))\ran^{N^B} = N^B(N^B+\can-1)\;.
\label{eqb11}
\ee
More importantly, $U(\can) \supset SO(\can)$ and the $\can(\can-1)/2$ generators 
of $SO(\can)$ are \cite{Ko-00},
\be
\barr{l}
SO(\can)\,:\, u^k_q(\ell , \ell)\;\;\mbox{with}\;\;k\;\;odd\;,\\
V^k_q(\ell_1 , \ell_2) = \l\{(-1)^{\ell_1 + \ell_2}
Y(\ell_1,\ell_2,k)\r\}^{1/2}\;
\l[\l(b^\dagger_{\ell_1}\tilde{b}_{\ell_2}\r)^k_q + Y(\ell_1,\ell_2,k)\,
\l(b^\dagger_{\ell_2}\tilde{b}_{\ell_1}\r)^k_q\r]\;;\\
Y(\ell_1,\ell_2,k)=(-1)^{k+1}\,\beta_{\ell_1} \beta_{\ell_2}\;.
\earr \label{eqb12}
\ee
Just as for fermion systems, the $SO(\can)$ is complimentary to the quasi-spin 
$SU^B_Q(1,1)$ and this follows from the relations (proved in \cite{Ko-00}),
\be
\barr{l}
4S^B_+S^B_- = C_2(U(\can))-{\hat n}^B -C_2(SO(\can))\;,\\
C_2(SO(\can))= \dis\sum_{\ell} C_2(SO(\can_\ell)) +
\dis\sum_{\ell_i < \ell_j} \dis\sum_{k}\;V^{k}\l(\ell_i, \ell_j\r)
\cdot V^{k}\l(\ell_i, \ell_j\r)\;; \\
\\
C_2(SO(\can_\ell)) = 2 \dis\sum_{k=\mbox{odd}}
u^{k}(\ell,\ell) \cdot u^{k}(\ell,\ell)\;; \\
\Longrightarrow \lan C_2(SO(\can))\ran^{N^B,\omega^B}= \omega^B(\omega^B +
\can-2)\;.  
\earr \label{eqb13}
\ee
In the last step we have used Eqs. (\ref{eqb6}) and (\ref{eqb11}). Thus,  the
irreps of  $SO(\can)$ are labeled by the symmetric irreps $\l[\omega^B\r]$ with
\be
\omega^B=N^B, N^B-2, \ldots, 0\;\mbox{or}\; 1\;.
\label{eqb13a}
\ee

\subsection{Seniority selection rules for one-body transition operators}

Given a general one-body operator
\be
\barr{l}
T^k_q= \dis\sum_{\ell_1, \ell_2}\epsilon^{k}_{\ell_1, \ell_2} \l(b^\dagger_{
\ell_1}\tilde{b}_{\ell_2}\r)^k_q \\
= \dis\sum_\ell \epsilon^{k}_{\ell, \ell}\l(b^\dagger_{\ell}\tilde{b}_{\ell}
\r)^k_q
+\dis\sum_{\ell_1 > \ell_2} \epsilon^{k}_{\ell_1, \ell_2} \l[
\l(b^\dagger_{\ell_1}\tilde{b}_{\ell_2}\r)^k_q +\dis\frac{\epsilon^{k}_{\ell_2,
\ell_1}}{\epsilon^{k}_{\ell_1, \ell_2}}\l(b^\dagger_{\ell_2}\tilde{b}_{\ell_1}
\r)^k_q\r]\;,
\earr \label{eqb14}
\ee
as $SO(N) \leftrightarrow SU_B(1,1)$, it should be clear from the generators in
Eq. (\ref{eqb12}) that the diagonal $\l(b^\dagger_{\ell}\tilde{b}_{\ell}
\r)^k_q$ part will be $SU^B_Q(1,1)$ scalar $T^0_0$ for $k$ odd and vector 
$T^1_0$ for $k$ even (except for $k=0$). Similarly, the off diagonal 
$$
\l[\l(b^\dagger_{\ell_1}\tilde{b}_{\ell_2}\r)^k_q +\dis\frac{\epsilon^{k}_{
\ell_2, \ell_1}}{\epsilon^{k}_{\ell_1, \ell_2}}\l(b^\dagger_{\ell_2}\tilde{b}_{
\ell_1}\r)^k_q\r]
$$
part will be 
\be
\dis\frac{\epsilon^{k}_{\ell_2, \ell_1}}{\epsilon^{k}_{\ell_1, \ell_2}}=
(-1)^{k+1} \beta_{\ell_1} \beta_{\ell_2} \rightarrow T^0_0\;,\;\;\;
\dis\frac{\epsilon^{k}_{\ell_2, \ell_1}}{\epsilon^{k}_{\ell_1, \ell_2}}=
(-1)^{k} \beta_{\ell_1} \beta_{\ell_2} \rightarrow T^1_0\;.
\label{eqb15}
\ee
Thus, the selection rules for the boson systems are similar to those for the
fermion systems. Results in Eqs. (\ref{eqb12}) and (\ref{eqb15}) together with a
condition for the seniority tensorial structure will allow us to write proper
forms for the EM operators in boson systems. Let us say that $S_+^B$ is given by
\be
S^B_+ = \dis\sum_{\ell} \dis\frac{\beta_\ell}{2} b^\dagger_\ell \cdot 
b^\dagger_\ell\;;\;\;\beta_l=+1\;\mbox{or}\;-1\;.
\label{eqb16}
\ee
If we impose the condition that the $T^{E,L=even}$ and $T^{M,L=odd}$ operators 
are $T^1_0$ and $T^0_0$ w.r.t. $SU_Q^B(1,1)$, just as the fermion operators 
are w.r.t. $SU_Q(2)$ (see Section III), then
\be
\barr{l}
T^{L}=\dis\sum_\ell \epsilon^{L}_{\ell, \ell}\l(b^\dagger_{\ell}\tilde{b}_{\ell}
\r)^L_q
+\dis\sum_{\ell_1 > \ell_2} \epsilon^{L}_{\ell_1, \ell_2} \l[
\l(b^\dagger_{\ell_1}\tilde{b}_{\ell_2}\r)^L_q +\beta_{\ell_1} \beta_{\ell_2}\,
\l(b^\dagger_{\ell_2}\tilde{b}_{\ell_1}\r)^L_q\r]\;;\\
\Rightarrow \\
T^{EL} \rightarrow T^1_0\;,\;\;\;T^{ML} \rightarrow T^0_0 \;.
\earr \label{eqb17}
\ee
Note that for $\ell_1 \neq \ell_2$, parity selection rule implies that
$(-1)^{\ell_1 + \ell_2}$ must be $+1$. Similarly, the parity changing
$T^{E,L=odd}$ and $T^{M,L=even}$ operators are,
\be
\barr{l}
T^{L}=\dis\sum_{\ell_1 > \ell_2} \epsilon^{L}_{\ell_1, \ell_2} 
\l[\l(b^\dagger_{\ell_1}\tilde{b}_{\ell_2}\r)^L_q - \beta_{\ell_1} 
\beta_{\ell_2}\,\l(b^\dagger_{\ell_2}\tilde{b}_{\ell_1}\r)^L_q\r]\;;\\
\Rightarrow \\
T^{EL} \rightarrow T^1_0\;,\;\;\;T^{ML} \rightarrow T^0_0 \;.
\earr \label{eqb18}
\ee
Note that for $\ell_1 \neq \ell_2$, parity selection rule implies that
$(-1)^{\ell_1 + \ell_2}$ must be $-1$ and therefore here we need orbits of
different parity as in $sp$ and $sdpf$ IBM's. On the other hand, if we impose
the condition that $T^{EL}$ is $T^0_0$ w.r.t. $SU_Q^B(1,1)$, then
\be
\barr{l}
T^{E,L=even}=\dis\sum_{\ell_1 > \ell_2} \epsilon^{L}_{\ell_1, \ell_2} 
\l[\l(b^\dagger_{\ell_1}\tilde{b}_{\ell_2}\r)^L_q -\beta_{\ell_1} 
\beta_{\ell_2}\,
\l(b^\dagger_{\ell_2}\tilde{b}_{\ell_1}\r)^L_q\r]\;;\;\;(-1)^{\ell_1 + 
\ell_2}=+1\;,\\
T^{E,L=odd}=\dis\sum_{\ell_1 > \ell_2} \epsilon^{L}_{\ell_1, \ell_2} 
\l[\l(b^\dagger_{\ell_1}\tilde{b}_{\ell_2}\r)^L_q +\beta_{\ell_1} 
\beta_{\ell_2}\,
\l(b^\dagger_{\ell_2}\tilde{b}_{\ell_1}\r)^L_q\r]\;;\;\;(-1)^{\ell_1 + 
\ell_2}=-1\\
\Rightarrow \\
T^{EL} \rightarrow T^0_0\;. 
\earr \label{eqb17ab}
\ee
Similarly, $T^{ML}$ can be chosen to be $T^1_0$ w.r.t. $SU_Q^B(1,1)$. Examples
for $sd$, $sp$, $sdg$ and $sdpf$ systems are discussed in Section VII.

\section{Number dependence of many particle matrix elements of $T^0_0$
and $T^1_0$ operators}

Applying Wigner-Ekart theorem for the many particle matrix elements in good
seniority states, number dependence of the matrix elements of $T^0_0$ and
$T^1_0$ operators is easily determined. For fermions one uses the $SU_Q(2)$
Wigner coefficients \cite{Edmonds} and for bosons $SU^B_Q(1,1)$ Wigner
coefficients \cite{Ui}. Results for fermions systems are given for example in
\cite{Talmi}. For completeness we will gives these here and also those for boson
systems. For fermions, using $SU_Q(2)$ algebra, we have
\be
\barr{l}
\lan m,v,\alpha \mid T^0_0 \mid m,v^\prime,\beta\ran =
\delta_{v,v^\prime} \lan v,v,\alpha \mid T^0_0 \mid v,v,\beta\ran\;, \\
\lan m,v,\alpha \mid T^1_0 \mid m,v,\beta\ran =
\dis\frac{\Omega-m}{\Omega-v}\,\lan v,v,\alpha \mid T^1_0 \mid v,v,\beta\ran\;,
\\
\lan m,v,\alpha \mid T^1_0 \mid m,v-2,\beta\ran = \dis\sqrt{
\dis\frac{(2\Omega-m-v+2)(m-v+2)}{4(\Omega-v+1)}}\,\lan v,v,\alpha \mid T^1_0 
\mid v,v-2,\beta\ran\;.\\
\earr \label{eqb19}
\ee
Similarly, for bosons, using $SU^B_Q(1,1)$ algebra (see \cite{Ui}), we have
\be
\barr{l}
\lan N^B,\omega^B,\alpha \mid T^0_0 \mid N^B,\omega^B,\beta\ran =
\lan \omega^B,\omega^B,\alpha \mid T^0_0 \mid \omega^B,\omega^B,\beta\ran\;, \\
\lan N^B,\omega^B,\alpha \mid T^1_0 \mid N^B,\omega^B,\beta\ran =
\dis\frac{\Omega^B+N^B}{\Omega^B+\omega^B}\,\lan \omega^B,\omega^B,\alpha 
\mid T^1_0 \mid \omega^B,\omega^B,\beta\ran\;,\\
\lan N^B,\omega^B,\alpha \mid T^1_0 \mid N^B,\omega^B-2,\beta\ran = \\
\dis\sqrt{\dis\frac{(2\Omega^B+N^B+\omega^B-2)(N^B-\omega^B+2)}{4(\Omega^B
+\omega^B-1)}}\,\lan \omega^B,\omega^B,\alpha \mid T^1_0 \mid \omega^B,
\omega^B-2,\beta\ran\;.\\
\earr \label{eqb20}
\ee
Note the well established $\Omega \rightarrow -\Omega$ symmetry between the
fermion and boson system formulas in  Eqs. (\ref{eqb19}) and (\ref{eqb20}); see
also \cite{KS-16,Ko-00}. Also, $T^1_0$ generates both $v (\omega^B) \rightarrow
v(\omega^B)$ and $v (\omega^B) \rightarrow v(\omega^B) \pm 2$ transitions while
$T^0_0$ only $v (\omega^B) \rightarrow v(\omega^B)$ transitions for fermion
(boson) systems. The later matrix elements are  independent of number of
particles.

In order to understand the variation of $B(EL)$ [similarly $B(ML$)] for fermion
systems, two numerical example are shown in Fig. 1. Firstly, considered an
electric multipole (of multipolarity $L$) transition between two states with
same $v$ value. Then, the $B(EL) \propto [\Omega-m)/(\Omega-v)]^2$ as seen from
the second equation in Eq. (\ref{eqb19}). Note that, with $\alpha_{j_i} =
(-1)^{\ell_i}$, the $T^{EL}$ operators are  $T^1_0$ w.r.t. $SU_Q(2)$. Assuming
$v=2$, variation of $B(EL)$ with particle number $m$ is shown for three
different values of $\Omega$ and $m$ varying from $2$ to $2 \Omega-2$. It is
clearly seen that $B(EL)$ decrease up to mid-shell and then again increases, i.e.
$B(EL)$ vs $m$ is an inverted parabola. This behavior is seen for example in Sn
isotopes  \cite{Jain1} as discussed further in Section VII.A. Going further,
assuming the ground $0^+$ and first excited $2^+$ states of a nucleus belong to
$v=0$ and $v=2$ respectively, $B(E2; 2^+ \rightarrow 0^+)$ variation with
particle number is calculated using the third equation in Eq. (\ref{eqb19})
giving $B(E2) \propto (2\Omega-m-v+2)(m-v+2)/4(\Omega-v+1)$. The variation of
$B(E2)$ is that it will increase up to mid-shell and then decrease; i.e. the
$B(E2; 2^+ \rightarrow 0^+)$ vs $m$ is a parabola. The behavior seen in Fig. 1b
is used to explain the variation in $B(E2)$'s in Sn isotopes \cite{Jain2} as
discussed further in Section VII.A.

In Fig. 2, variation of $B(EL)$ for boson systems with boson number assuming
$\omega^B=2$ for both $\omega^B \rightarrow \omega^B$ and $\omega^B \rightarrow
\omega^B-2$ transitions are shown by employing the last two equations in Eq.
(\ref{eqb20}).  The $B(EL)$ values increase with $N^B$ and this variation is
quite different from the variation seen in Fig. 1 for fermion systems. The
increase in $B(E2)$'s with increase in $N^B$ is seen in Ru isotopes \cite{spdf}.

\begin{figure*}
\includegraphics[width=6cm]{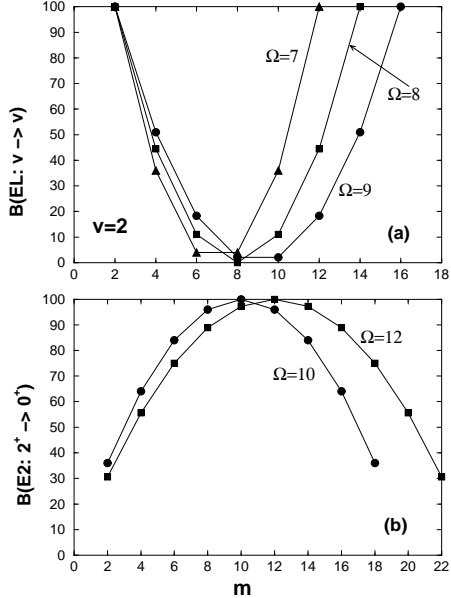}

\caption{(a) Variation of $B(EL)$ with particle number $m$ for three different
values of $\Omega$ and seniority $v=2$ for $v \rightarrow v$ transitions. (b)
Variation of $B(E2; v=2,2^+ \rightarrow v=0,0^+)$ with particle number. Results
are for fermion systems and they are obtained by applying the last two equations
in Eq. (\ref{eqb19}). The $B(EL)$ and $B(E2)$ values are scaled such that the
maximum value is 100 and they are not in any units.}  

\label{be-fermi}
\end{figure*}

\begin{figure*}
\includegraphics[width=6cm]{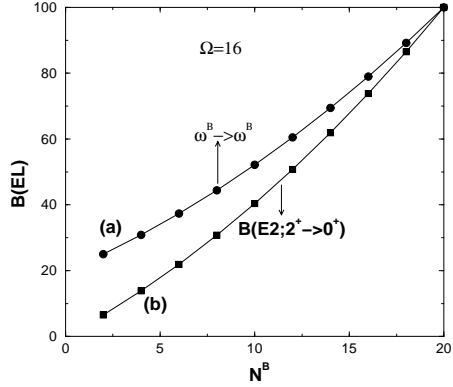}

\caption{(a) Variation of $B(EL)$ with particle number $N^B$ for $\Omega=16$ 
and seniority $\omega^B=2$ for $\omega^B \rightarrow \omega^B$ transitions. (b)
Variation of $B(E2; \omega^B=2,2^+ \rightarrow \omega^B=0,0^+)$ with particle 
number $N^B$. Results are for boson systems and they are  obtained by applying
the last two equations in Eq. (\ref{eqb20}). The $B(EL)$ and $B(E2)$ values are
scaled such that the maximum value is 100 and they are not in any units.}  

\label{be-bose}
\end{figure*}

\section{Applications}

\subsection{Shell model applications}

First examples for the goodness of generalized seniority in nuclei are Sn
isotopes.  Note that for Sn isotopes the valence nucleons are neutrons with Z=50
a magic number. From Eq. (\ref{eq6}) it is easy to see that the spacing between
the first $2^+$ state (it will have $v=2$) and the ground state $0^+$ (it will
have $v=0$) will be independent of $m$, i.e. the spacing should be same for all
Sn isotopes and this is well verified by experimental data \cite{Talmi}. Going
beyond this, recently $B(E2; 2^+_1 \rightarrow 0^+_1)$ data for $^{104}$Sn to
$^{130}$Sn are analyzed using the results in Eq. (\ref{eqb19}), i.e. the results
in Fig. 1b. Data shows a dip at $^{116}$Sn and they are close to adding two
displaced parabolas; see Fig. 1 in \cite{Jain2}. This is understood by employing
$^0g_{7/2}$, $^1d_{5/2}$, $^1d_{3/2}$ and $^2s_{1/2}$ orbits for neutrons in
$^{104}$Sn to $^{116}$Sn with $\Omega=10$ and $^{100}$Sn core. Similarly,
$^1d_{5/2}$, $^1d_{3/2}$, $^2s_{1/2}$ and $^0h_{11/2}$ orbits   with $\Omega=12$
and $^{108}$Sn core for $^{116}$Sn to $^{130}$Sn. Then the $B(E2)$ vs $m$
structure follows from Fig. 1b by shifting appropriately the centers of the two
parabolas in the figure and defining properly the beginning and end points. It
is also shown in \cite{Jain2} that shell model calculations with an appropriate
effective interaction in the above orbital spaces reproduce the results from the
simple formulas given by seniority description and the experimental data.

In addition to $B(E2; 2^+ \rightarrow 0^+)$ data, there is now good data
available for $B(E2)$'s and $B(E1)$'s for some high-spin isomer states in even
Sn isotopes. These are: $B(E2; 10^+\rightarrow 8^+)$ data for $^{116}$Sn to
$^{130}$Sn and $B(E2; 15^- \rightarrow 13^-)$ for $^{120}$Sn to $^{128}$Sn and
$B(E1; 13^- \rightarrow 12^+)$ in $^{120}$Sn to $^{126}$Sn. The states $10^+$
and $8^+$ are interpreted to be $v=2$ states while $15^-$, $13^-$ and $12^+$
are $v=4$ states. Therefore, all these transitions are $v \rightarrow v$
transitions and the their variation with $m$ will be as shown in Fig. 1a.
This is well verified by data \cite{Jain1} by assuming that the active sp
orbits are $^0h_{11/2}$, $^1d_{3/2}$ and $^2s_{1/2}$ with $\Omega=9$ (see also
Fig. 1a with $\Omega=9$). The results with $\Omega=8$ and $\Omega=7$ obtained by
dropping  $^2s_{1/2}$ and $^1d_{3/2}$ orbits respectively, are not in good
accord with data.

In summary, both the $B(E2; 2^+ \rightarrow 0^+)$ data and the $B(E2)$  and
$B(E1)$ data for high-spin isomer states are explained by assuming goodness of
generalized seniority with the choice $\beta_j=(-1)^{\ell_j}$ but with 
effective $\Omega$ values. Although the sp orbits (and hence $\Omega$ values)
used are different for the low-lying levels and the high-spin isomer states, the
good agreements between data and  effective generalized seniority description on
one hand and the correlation coefficients presented in Section IV on the other
show that for Sn isotopes generalized seniority is possibly an 
\lq{emergent symmetry}\rq.

In addition to even Sn isotopes, $B(E2)$ data for $10^+$ isomers in $N=82$
isotones (A=148-162), $12^+$ isomers in Pb isotopes (A=176-198) and also
high-spin isomers in odd-A Sn isotopes, N=82 isotones and Pb isotopes are
analyzed, though the data is sparse, using the results in Fig. 1 and Eq.
(\ref{eqb19}) \cite{Jain3}.

\begin{figure*}
\includegraphics[width=5in]{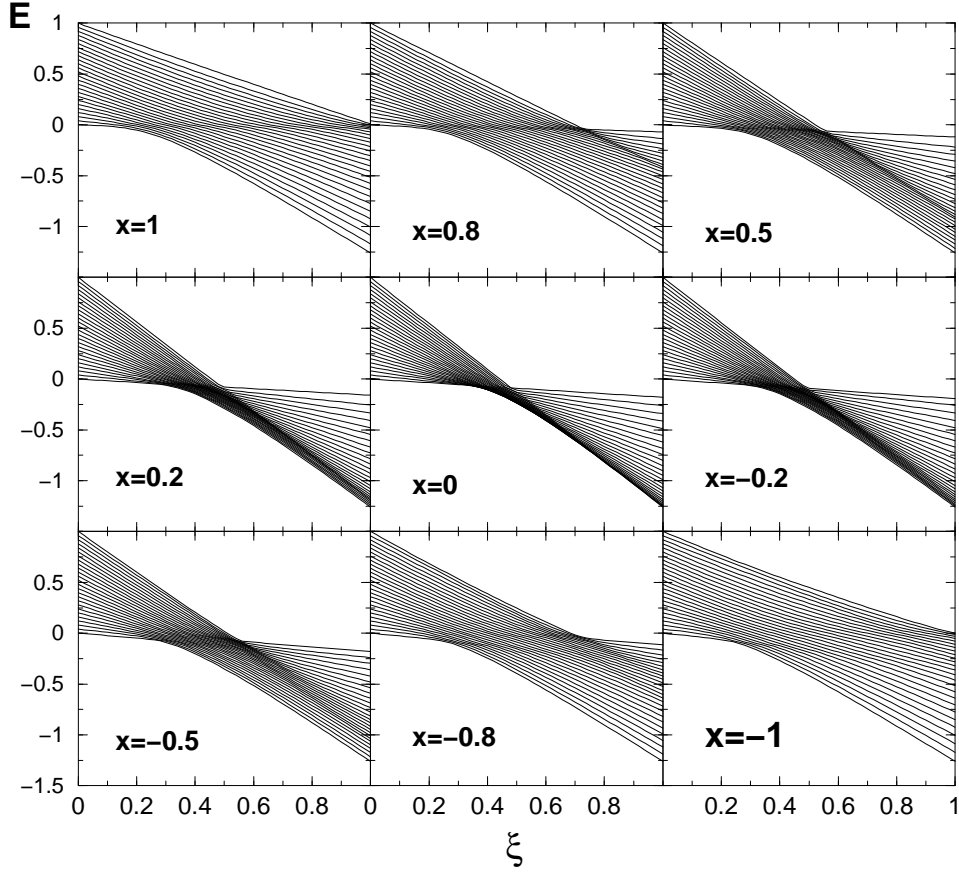}

\caption{Energy spectra for 50 bosons and $(\omega^B_{sd},\omega^B_g)=(0,0)$ in
$sdg$IBM with the Hamiltonian $H_{sdg}= [(1-\xi)/N^B]\,\hat{n}_g +
[(\xi/(N^B)^2]\,[4(S_+^{sd}+xS_+^g)(S_-^{sd}+xS_-^g) -N^B(N^B+13)]$ where
$S_+^{sd}$ is the $S_+$ operator for the $sd$ boson system and $S_+^g$ for the
$g$ boson system. In each panel, energy spectra are shown as a function of the
parameter $\xi$ taking values from $0$ to $1$. Results are shown in the figures
for $x=1$, $0.8$, $0.5$, $0.2$, $0$, $-0.2$, $-0.5$, $-0.8$ and $-1$. Note that
$x=1$ and $-1$ correspond to the two $SU(1,1)$ algebras in the model. In the
figures, energies are not in any units.}  

\label{qpt}
\end{figure*}

\subsection{Interacting boson model applications}

Turning to the interacting boson models, let us first  consider the $SO(6)$
limit of  $sd$IBM. Then, we have $U(6) \supset SO(6)$ and the complimentary
$SU(1,1)$ algebra corresponds to the $sd$ pair $S_+=s^\dagger s^\dagger\pm
d^\dagger \cdot d^\dagger$. Arima and Iachello \cite{Iac-87} used the choice 
$S_+= s^\dagger s^\dagger - d^\dagger \cdot d^\dagger$. The corresponding
$SU(1,1)$ we denote as $SU^{-}(1,1)$. Similarly, the $SU(1,1)$ with  $S_+=
s^\dagger s^\dagger + d^\dagger \cdot d^\dagger$ is denoted by $SU^+(1,1)$.
Corresponding to the two $SU(1,1)$ algebras, there will be two $SO(6)$ algebras
as pointed out first in \cite{so6-1}. Their significance is seen in quantum
chaos studies \cite{so6-2,so6-3}. For illustration, let us consider the
tensorial structure of the $E2$ operator. Following the discussion in Section V,
the $E2$ transition operator will be $T^0_0$ w.r.t. $SU^-(1,1)$ if we choose
$T^{E2} = \alpha (s^\dagger \tilde{d} + d^\dagger \tilde{s})^2_\mu$ where
$\alpha$ is a constant.  This is the choice made in \cite{Iac-87} and this
operator will not change the seniority quantum number (called $\sigma$ in
\cite{Iac-87}) defining the irreps of $SO(6)$ that is complimentary to
$SU^-(1,1)$. However, if we demand that the $T^{E2}$ operator should be $T^1_0$
w.r.t. $SU^-(1,1)$, then  we have $T^{E2}=\alpha_1 (d^\dagger \tilde{d})^2_\mu +
i \alpha_2  (s^\dagger \tilde{d} - d^\dagger \tilde{s})^2_\mu$. This operator
will have both $\sigma \rightarrow \sigma$ and $\sigma \rightarrow \sigma \pm 2$
transitions. On the other hand,  $T^{E2}=\alpha_1 (d^\dagger \tilde{d})^2_\mu +
\alpha_2  (s^\dagger \tilde{d} + d^\dagger \tilde{s})^2_\mu$ will be a mixture
of $T^0_0$ and $T^1_0$ operators.

In the second example we will consider the $sp$ boson model, also called vibron
model with applications to diatomic molecules \cite{so4-1} and two-body 
clusters in nuclei \cite{CLUST}. Just as in $sd$IBM, here we have $U(4) \supset
SO(4)$ and there will be two $SO(4)$ algebras with $S_+=s^\dagger s^\dagger +
\beta p^\dagger \cdot p^\dagger$; $\beta=\pm 1$. The general form of the $E1$
operator (to lowest order) in this model is $T^{E1}=\epsilon_{sp} \l(s^\dagger
\tilde{p} \pm p^\dagger \tilde{s}\r)^1_\mu$. With $SU^+(1,1)$ defined by 
$S_+=s^\dagger s^\dagger + p^\dagger \cdot p^\dagger$,  from Eq. (\ref{eqb18})
we see that $T^{E1}=i \epsilon \l(s^\dagger \tilde{p} - p^\dagger 
\tilde{s}\r)^1_\mu$ will be $T^1_0$ w.r.t. $SU^+(1,1)$. Similarly, with
$SU^{-}(1,1)$ defined by  $S_+=s^\dagger s^\dagger - p^\dagger \cdot
p^\dagger$,  from Eq. (\ref{eqb17}) we see that $T^{E1}=\epsilon_{sp}
\l(s^\dagger \tilde{p}  + p^\dagger  \tilde{s}\r)^1_\mu$ will be $T^1_0$ w.r.t.
$SU^-(1,1)$. If the definitions of $T^{E1}$ are interchanged, then they will be
$T^0_0$ w.r.t. the corresponding $SU(1,1)$ algebras. These results are described
and applied in \cite{so4-1,so4-2,so4-3}.        

In the third example we will consider the $sdg$ interacting boson model
\cite{sdg-1} and there is new interest in this model in the context of quantum
phase transitions (QPT) \cite{sdg-2}. With $s$, $d$ and $g$ bosons, the
generalized pair operator here is $S_+=s^\dagger s^\dagger  \pm d^\dagger \cdot
d^\dagger \pm g^\dagger \cdot g^\dagger$ giving four $SU^{+,\pm,\pm}(1,1)$
algebras and the  corresponding $SO^{+,\pm,\pm}(15)$ algebras in $U(15) \supset
SO(15)$;  the superscripts in $SU^{+,\pm,\pm}(1,1)$ and similarly in $SO(15)$ 
are the signs of the $s$ , $d$ and $g$ pair operators in $S_+$. In QPT studies,
VanIsacker et al have chosen \cite{sdg-2} the operators $(s^\dagger \tilde{d} +
d^\dagger \tilde{s})^2_\mu$ and  $(s^\dagger \tilde{g} + g^\dagger
\tilde{s})^4_\mu$ to be $SO(15)$ scalars. Then, from Eq. (\ref{eqb15}) it is
seen that the $SO(15)$ will correspond to the $SU^{+,-,-}(1,1)$ algebra with
$H_p=S_+S_-$ where $S_+=s^\dagger s^\dagger  - d^\dagger \cdot d^\dagger -
g^\dagger \cdot  g^\dagger$. Note that here the $sd$-part is same as the
one used by Arima and Iachello (see the $sd$IBM discussion above). In another 
recent study, the $E2$ operator in $sdg$IBM was chosen to be \cite{sdg-3}
$$
T^{E2}=\alpha_1 (d^\dagger \tilde{d})^2_\mu + \alpha_2 (g^\dagger 
\tilde{g})^2_\mu + \alpha_3 (s^\dagger \tilde{d} + d^\dagger \tilde{s})^2_\mu
+ \alpha_4 (d^\dagger \tilde{g} + g^\dagger \tilde{d})^2_\mu\;.
$$
With respect to the $SU^{+,-,-}(1,1)$ above, this operator will be a mixture
of $T^0_0$ and $T^1_0$ operators. However, w.r.t. $SU^{+,+,+}(1,1)$, it will be
a pure $T^1_0$ operator. It should be clear that with different choices of 
$SU(1,1)$ algebras (there are four of them), the QPT results for transition to
rotational $SU(3)$ limit in $sdg$IBM will be different. It is important to 
investigate this going beyond the results presented in \cite{sdg-2}.

In the final example, let us consider the $sdpf$ model \cite{Kusnezov} applied
recently with good success in describing $E1$ strength distributions in Nd, Sm,
Gd and Dy isotopes \cite{Iac-15} and also spectroscopic properties (spectra,
and  $E2$ and $E1$ strengths) of even-even $^{98-110}$Ru isotopes \cite{spdf}.
Note that the parities of the $p$ and $f$ orbit are negative. In $sdpf$IBM,
following the results in Section V, there will be eight generalized pairs $S_+$
and the algebra complimentary to the $SU(1,1)$ is $SO(16)$ in $U(16) \supset
SO(16)$. Keeping the $SO(6)$ pair structure, as chosen by Arima and Iachello, of
$sd$IBM intact we will have four $S_+$ pairs, $S_+= s^\dagger s^\dagger -
d^\dagger \cdot d^\dagger \pm p^\dagger \cdot p^\dagger \pm f^\dagger \cdot
f^\dagger$ giving $SU^{+,-,\pm,\pm}(1,1)$ and correspondingly four
$SO^{+,-,\pm,\pm}(16)$ algebras. For each of the four choices, one can write
down the $T^{E2}$ and $T^{E1}$ operators that transform as $T^0_0$ or $T^1_0$
w.r.t. $SU(1,1)$. In \cite{Kusnezov}, $SU^{+,-,-,-}(1,1)$ is employed. Then, the
$E2$ and $E1$ operators employed in \cite{spdf,Iac-15,Kusnezov} will be mixture
of $T^0_0$ and $T^1_0$ w.r.t. to $SU^{+,-,-,-}(1,1)$. For example, the $E1$
operator used is,
\be
T^{E1}=\alpha_{sp}\l(s^\dagger \tilde{p} + p^\dagger \tilde{s}\r)^1_\mu +
\alpha_{pd}\l(p^\dagger \tilde{d} + d^\dagger \tilde{p}\r)^1_\mu +
\alpha_{df}\l(d^\dagger \tilde{f} + f^\dagger \tilde{d}\r)^1_\mu\;.
\label{eq-e1op}
\ee
The first term in the operator will be $T^1_0$ and the remaining two terms will
be $T^0_0$ w.r.t. $SU^{+,-,-,-}(1,1)$. However if we use $ \alpha_{sp} 
\l(s^\dagger \tilde{p} - p^\dagger \tilde{s}\r)^1_\mu$  in the above, then the
whole operator will be $T^0_0$. It will be interesting to employ the
$H_p=S_+S_-$ with $S_+$ given above (there will be four choices) in the analysis
made in \cite{spdf} and confront the data. 

\section{Conclusions}

In this article an attempt is made to bring focus, bringing all known and new
results to one place, to  multiple multi-orbit pairing algebras in $j-j$
coupling shell model for identical nucleons and similarly, for identical boson
systems described by multi-orbit interacting boson models such as $sd$, $sp$,
$sdg$ and $sdpf$ IBM's.  The relationship between quasi-spin tensorial nature of
one-body transition operators and the phase choices in the multi-orbit pair
creation operator is derived for both identical fermion (described by shell
model) and boson (described by interacting boson model) systems.  These results
are presented in Sections II and III for fermion systems and V for boson
systems. As pointed out in these sections, some of the results here are known
before for some special situations. In Section IV, results for the correlation
coefficient between the pairing operator with different choices for phases in
the generalized pair creation operator and realistic effective interactions are
presented. It is found that the choice  advocated by AM \cite{Arvieu} gives 
maximum correlation though its absolute value is no more than 0.3. Particle
number variation in electromagnetic transition strengths is discussed in Section
VI. 

Applications of multiple pairing algebras are briefly discussed in Section VII.
As discussed in Section VII.A, drawing from the recent analysis by
Maheswari and Jain \cite{Jain1,Jain2}, shell model generalized seniority with
phase choice advocated by AM appear to describe $B(E2)$ and $B(E1)$ data  in Sn
isotopes both for low-lying states and high-spin isomeric states. Though
deviations from the results obtained using AM choice is a signature for 
multiple multi-orbit pairing algebras, direct experimental evidence for the 
multiple pairing algebras is not yet available. 

Turning to interacting boson model description of collective states, imposing
specific tensorial structure, with respect to pairing $SU(1,1)$ algebras, is
possible as discussed with various examples in Section VII.B. It will be
interesting to derive results  for B(E2)'s (say in $sdg$ and $sdpf$ IBM's) and
B(E1)'s (in $sdpf$ IBM) with fixed tensorial structure for the transition
operator but with wavefunctions that correspond to different $SU(1,1)$ algebras.
Such an exercise was carried out before for $sd$IBM \cite{so6-1}. Also, with
recent interest in $sdg$ \cite{sdg-2} and $sdpf$ \cite{Iac-15,spdf} IBM's, it
will be interesting to study quantum phase transitions and order-chaos
transitions in these models, in a systematic way, employing Hamiltonians that
interpolate the different pairing  algebras in these models. Such studies for
the simpler $sd$ and $sp$ IBM's are available; see for example
\cite{so6-2,so6-3,casten}. Construction of the Hamiltonian matrix for the
interpolating Hamiltonians is straightforward as described briefly in
Appendix-A. As an example, results for the spectra for a $sdg$IBM system are 
shown in Fig. 3.

Finally, going beyond multiple pairing algebras for identical fermion or boson
systems, there are also multiple pairing algebras for fermions and bosons
carrying internal degrees of freedom such as isospin and spin. Though these are
identified \cite{KS-16,Ko-00,so8}, they are not studied in any detail till now.
Similarly, there are multiple rotational $SU(3)$ algebras both in shell model 
and IBM's as discussed  with some specific examples in
\cite{Parikh,Iac-87,casten,kota-sdg}. A systematic study of these multiple
extended pairing algebras with internal degrees of freedom and multiple $SU(3)$
algebras will be the topics of future papers.

\acknowledgments

Thanks are due to A.K. Jain for very useful correspondence. Thanks are also due
to Feng Pan for useful discussions during the visit to Dalian.

\renewcommand{\theequation}{A-\arabic{equation}}
\setcounter{equation}{0}   

\section*{APPENDIX A}

Let us consider identical bosons in $r$ number of $\ell$ orbits. Now, given the
general pairing Hamiltonian 
\be
\barr{l}
H^G_p=\dis\sum_{\ell} \epsilon_\ell \hat{n}^B_\ell + S^B_+ S^B_-\;; \\
S^B_+ = \dis\sum_{\ell} x_\ell S^B_+(\ell)\;,\;S^B_+(\ell)=
\spin b^\dagger_\ell \cdot b^\dagger_\ell\;,
\earr \label{eqap1}
\ee
it will interpolate $U(\can) \supset SO(\can) \supset \sum_\ell SO(\can_\ell) 
\oplus$ pairing algebra and $U(\can) \supset \sum_\ell [U(\can_\ell) \supset 
SO(\can_\ell)] \oplus$ pairing algebra for arbitrary values of $x_\ell$'s and 
$\epsilon_\ell$'s. Matrix representation for the Hamiltonian $H^G_p$ is easy 
to construct by choosing the basis
\be
\Phi=\l.\l|N^B_{\ell_1}, \omega^B_{\ell_1}, \alpha_{\ell_1}; N^B_{\ell_2}, 
\omega^B_{\ell_2}, \alpha_{\ell_2}; \ldots N^B_{\ell_r}, \omega^B_{\ell_r}, 
\alpha_{\ell_r}\r.\ran
\label{eqap2}
\ee      
where $\alpha_{\ell_i}$ are additional labels required for complete
specification of the basis states (they play no role in the present discussion)
and total number of bosons $N^B=\sum_\ell N^B_\ell$. The first term (one-body
term) in $H^G_p$ is diagonal in the $\Phi$ basis giving simply $\sum_\ell
\epsilon_\ell N^B_\ell$. The second term can be written as
\be
S^B_+ S^B_- = \l[\dis\sum_\ell (x_\ell)^2 S^B_+(\ell) S^B_-(\ell)\r] +
\l[\dis\sum_{\ell_i \neq \ell_j} x_{\ell_i} x_{\ell_j} S^B_+(\ell_i) 
S^B_-(\ell_j)\r]\;.
\label{eqap3}
\ee
In the basis $\Phi$, the first term is diagonal and its matrix elements follow 
directly from Eq. (\ref{eqb6}) and it is the second term that mixes the basis 
states $\Phi$'s. The mixing matrix elements follow from,
\be
\barr{l}
S_+^B(\ell_i)S^B_-(\ell_j) \l.\l|N^B_{\ell_i}, \omega^B_{\ell_i}
\alpha_{\ell_i}; N^B_{\ell_j}, \omega^B_{\ell_j}, \alpha_{\ell_j}\r.\ran
\\
=\dis\frac{1}{4}\dis\sqrt{\l(N^B_{\ell_j}-\omega^B_{\ell_j}\r)\l(
2\Omega^B_{\ell_j} + N^B_{\ell_j} +\omega^B_{\ell_j} -2\r)
\l(N^B_{\ell_i}-\omega^B_{\ell_i} +2\r)\l(2\Omega^B_{\ell_i} + N^B_{\ell_i} +
\omega^B_{\ell_i}\r)} \\
\l.\l|N^B_{\ell_i}+2, \omega^B_{\ell_i}
\alpha_{\ell_i}; N^B_{\ell_j}-2, \omega^B_{\ell_j}, \alpha_{\ell_j}\r.\ran \;.
\earr \label{eqap4}
\ee
It is important to note that the action of $H^G_p$ on the basis states $\Phi$ will
not change the $\omega^B_\ell$ quantum numbers. For boson numbers not large, it
is easy to apply Eqs. (\ref{eqap3}) and (\ref{eqap4}) and construct the $H^G_P$
matrices. It is easy to extend the above formulation to fermion systems and also
for the situation where two or more orbits are combined to a larger orbit. The
later, for example for $sdg$IBM gives $U(15) \supset SO(15) \supset SO_{sd}(6)
\oplus SO_g(9)$ and $U(15) \supset [U(6) \supset SO(6)] \oplus [U(9) \supset
SO(9)]$ interpolation [similarly with $SO_{dg}(14)$ and $SO_{sg}(10)$ algebras].
Finally, it is also possible to use the exact solution for the generalized
pairing Hamiltonians as given in \cite{Pan}  which is more useful for large
boson numbers.

\ed